\documentclass[showpacs,aps,prl,superscriptaddress,reprint]{revtex4-1}
\usepackage{graphicx,amsmath,amssymb,epsfig,color,lineno}
\usepackage{verbatim} 
\usepackage{xspace,afterpage,natbib,psfrag}
\input{babarsym}

\newcommand{\BABARPubYear}    {12}
\newcommand{\BABARPubNumber}  {012}
\newcommand{\SLACPubNumber} {15028}

\newcommand{\dss}{\ensuremath{D^{**}}\xspace}
\newcommand{\ds}{\ensuremath{D^{(*)}}\xspace}

\newcommand{\dspizl}{\ensuremath{D^{(*)}\piz\ell}\xspace}
\newcommand{\Btag}{\ensuremath{B_{\rm tag}}\xspace}
\newcommand{\elltau}{\ensuremath{(\ell/\tau)}\xspace}

\newcommand{\BDtaunu}{\ensuremath{\Bb \rightarrow D\tau^-\nutb}\xspace}
\newcommand{\BDstaunu}{\ensuremath{\Bb \rightarrow D^*\tau^-\nutb}\xspace}
\newcommand{\BDxtaunu}{\ensuremath{\Bb \rightarrow D^{(*)}\tau^-\nutb}\xspace}

\newcommand{\BDxlnu}{\ensuremath{\Bb \rightarrow D^{(*)}\ell^-\nub_{\ell}}\xspace}
\newcommand{\BDssltnu}{\ensuremath{B \rightarrow D^{**}(\ell/\tau)\nu}\xspace}
\newcommand{\BDsslnu}{\ensuremath{B \rightarrow D^{**}\ell\nu}\xspace}
\newcommand{\BDsstaunu}{\ensuremath{B \rightarrow D^{**}\tau\nu}\xspace}

\newcommand{\Dsslnu}{\ensuremath{D^{**}(\ell/\tau)\nu}\xspace}

\newcommand{\RDx}{\ensuremath{{\cal R}(D^{(*)})}}
\newcommand{\RDs}{\ensuremath{{\cal R}(D^*)}}
\newcommand{\RD}{\ensuremath{{\cal R}(D)}}
\newcommand{\RDz}{\ensuremath{{\cal R}(\Dz)} }
\newcommand{\RDp}{\ensuremath{{\cal R}(\Dp)} }
\newcommand{\RDstarz}{\ensuremath{{\cal R}(\Dstarz)}}
\newcommand{\RDstarp}{\ensuremath{{\cal R}(\Dstarp)}}

\newcommand{\mmiss}{\ensuremath{m_{\rm miss}^2}\xspace}
\newcommand{\mES}{\ensuremath{m_{ES}}\xspace}
\newcommand{\eextra}{\ensuremath{E_{\rm extra}}\xspace}
\newcommand{\pstarl}{\ensuremath{|\boldsymbol{p}^*_\ell|}\xspace}

\newcommand{\tanB}{\ensuremath{{\rm tan}\beta}\xspace}
\newcommand{\mH}{\ensuremath{m_{H^+}}\xspace}
\newcommand{\tBmH}{\ensuremath{\tanB/\mH}\xspace}

\begin{document}

\preprint{\babar-PUB-\BABARPubYear/\BABARPubNumber} 
\preprint{SLAC-PUB-\SLACPubNumber} 
  
\begin{flushleft}
hep-ex/1205.5442\\
\babar-PUB-\BABARPubYear/\BABARPubNumber\\
SLAC-PUB-\SLACPubNumber\\
\end{flushleft}
  
\title{ Evidence for an excess of {\boldmath \BDxtaunu} decays} \vspace{0.3in}
%
\author{J.~P.~Lees}
\author{V.~Poireau}
\author{V.~Tisserand}
\affiliation{Laboratoire d'Annecy-le-Vieux de Physique des Particules (LAPP), Universit\'e de Savoie, CNRS/IN2P3,  F-74941 Annecy-Le-Vieux, France}
\author{J.~Garra~Tico}
\author{E.~Grauges}
\affiliation{Universitat de Barcelona, Facultat de Fisica, Departament ECM, E-08028 Barcelona, Spain }
\author{A.~Palano$^{ab}$ }
\affiliation{INFN Sezione di Bari$^{a}$; Dipartimento di Fisica, Universit\`a di Bari$^{b}$, I-70126 Bari, Italy }
\author{G.~Eigen}
\author{B.~Stugu}
\affiliation{University of Bergen, Institute of Physics, N-5007 Bergen, Norway }
\author{D.~N.~Brown}
\author{L.~T.~Kerth}
\author{Yu.~G.~Kolomensky}
\author{G.~Lynch}
\affiliation{Lawrence Berkeley National Laboratory and University of California, Berkeley, California 94720, USA }
% ...
\collaboration{The \babar\ Collaboration}
\noaffiliation

\date{August 14, 2012}

\begin{abstract}
Based on the full \babar\ data sample, we report improved measurements of the ratios
$ \RDx = {\cal B}(\BDxtaunu)/{\cal B}(\BDxlnu)$, where $\ell$ is either $e$  or $\mu$.
These ratios are sensitive to new physics contributions in the form of a charged Higgs boson. 
We measure $\RD  = 0.440\pm 0.058\pm 0.042$ and $\RDs = 0.332\pm 0.024\pm 0.018$,
which exceed the standard model expectations by $2.0\sigma$ and $2.7\sigma$, respectively. 
Taken together, our results disagree with these expectations at 
the $3.4\sigma$ level. This excess cannot be explained by a charged Higgs boson in the
type II two-Higgs-doublet model.
\end{abstract}

\maketitle

In the standard model (SM), semileptonic decays of $B$ mesons are well-understood 
processes mediated by a $W$ boson~\cite{Heiliger:1989yp,Korner:1989qb,Hwang:2000xe}. 
Decays involving the higher mass $\tau$ lepton are sensitive to 
additional amplitudes, such as those involving an intermediate charged Higgs 
boson~\cite{Tanaka:1994ay,Itoh:2004ye,Nierste:2008qe,Tanaka:2010se,Fajfer:2012vx},
and offer an excellent opportunity to search for this and other non-SM contributions.

Our understanding of exclusive semileptonic decays has greatly improved  over the past two decades, 
thanks to the development of heavy-quark effective theory and precise
measurements of \BDxlnu~\cite{convention} at the $B$ factories~\cite{Antonelli:2009ws,Nakamura:2010zzi}.
SM expectations for the relative rates $ \RDx = {\cal B}(\BDxtaunu)/{\cal B}(\BDxlnu)$ have 
less than 6\% uncertainty~\cite{Fajfer:2012vx}.
Calculations~\cite{Tanaka:1994ay,Itoh:2004ye,Nierste:2008qe,Tanaka:2010se,Fajfer:2012vx} 
based on two-Higgs-doublet models (2HDM) predict a
substantial impact on the ratio \RD and a smaller effect on \RDs.
The ratios \RD\ and \RDs\ are independent of the Cabibbo-Kobayashi-Maskawa element 
$|V_{cb}|$ and also, to a large extent, of the parameterization of the hadronic matrix elements.

The decay \BDstaunu was first observed in 2007 by the Belle Collaboration~\cite{Matyja:2007kt}.
Since then, both \babar\ and Belle have published improved measurements
and have found evidence for \BDtaunu decays~\cite{Aubert:2007dsa,Adachi:2009qg,Bozek:2010xy}. 
Although the measured values for \RD\ and \RDs\ have consistently exceeded the SM expectations, 
the significance of the excess has remained low due to the large statistical uncertainties. 

This analysis is an update of an earlier \babar\ measurement~\cite{Aubert:2007dsa}. It is based on the full 
\babar\ data sample and includes improvements to the event reconstruction that increase the signal 
efficiency by more than a factor of 3. 

We analyze data recorded with the \babar\ detector~\cite{Aubert:2001tu} at a center-of-mass (c.m.) energy
of 10.58 \gev, corresponding to the mass of the $\FourS$ resonance, which
decays almost exclusively to \BB\ pairs. The data sample comprises an integrated luminosity 
of 426\invfb , and contains
$471\times 10^6$ \BB\ pairs. An additional sample of $40\invfb$, taken at a c.m. energy 40 \mev 
below the \FourS resonance (off-peak data), is used to study continuum background from $\epem\to 
f\overline{f}(\gamma)$ pair production with $f=u,d,s,c,\tau$.  

We choose to reconstruct only the purely leptonic decays of the $\tau$ lepton,
$\tau^- \to e^- \nueb \nut$ and $\tau^- \to \mu^- \numb\nut$, so that \BDxtaunu (signal)
and \BDxlnu (normalization) events are identified by the same particles in the final
state. This leads to the cancellation of various sources of uncertainty 
in the ratios \RDx. Events corresponding to $\FourS\to\BB$ decays 
are selected by reconstructing the hadronic 
decay of one of the $B$ mesons (\Btag), a \ds meson and a lepton ($e$ or $\mu$).
Signal and normalization yields are extracted from a fit to the spectra of two variables: the invariant mass 
of the undetected particles $\mmiss=p^2_{\rm miss}=(p_{\epem} - p_{\rm tag} - p_{D^{(*)}} - p_{\ell})^2$ 
(where $p$ denotes the four-momenta of the colliding beams, the \Btag, the \ds, and the charged 
lepton) and the lepton three-momentum in the $B$ rest frame \pstarl.
The \mmiss distribution of decays with a single missing neutrino peaks at zero, whereas signal events,
which have three missing neutrinos,
have a broad \mmiss distribution that extends to about $9 \gev ^2$.
The observed lepton in signal events is a secondary particle from the $\tau$ decay, 
so its \pstarl spectrum is softer than for normalization events.

The \Btag reconstruction has been greatly improved with respect to previous analyses~\cite{Aubert:2003zw}. 
We now reconstruct \Btag candidates in 1680 final states.
We look for decays of the type $\Btag \to S X^{\pm}$, where $S$ refers to a 
seed meson (\Dz, \Dstarz, \Dp, \Dstarp, \Ds, \Dss, or $\jpsi$) reconstructed in 56 different decay modes,
and $X^{\pm}$ is a charged state decaying to up to five hadrons (\pipm, \Kpm, $\pi^0$, and \KS). 
Two kinematic variables are used to select \Btag candidates: 
$\mES=\sqrt{E^{2}_{\rm beam} - \boldsymbol{p}^{2}_{\rm tag}}$ and 
$\Delta E = E_{\rm tag} - E_{\rm beam}$.  Here $\boldsymbol{p}_{\rm tag}$ and $E_{\rm tag}$ 
refer to the c.m. momentum and energy of the $B_{\rm tag}$, and  $E_{\rm beam}$ is the c.m.~energy 
of a single beam particle.  For correctly reconstructed 
$B$ decays, the $m_{ES}$ distribution is centered at the $B$-meson mass with a resolution 
of 2.5 \mev, while $\Delta E$ is centered at zero with a resolution of 
18\mev. We require 
$m_{ES} > 5.27 \gev$ and $|\Delta E| <0.072 \gev$. 

We combine each \Btag candidate with a $D^{(*)}$ meson candidate and a charged lepton $\ell$. 
Events with additional charged particles are rejected. The laboratory momentum of the electron 
or muon is required to exceed 300\mev or 200\mev, respectively.
$D$ decays are reconstructed in the following decay modes:
$\Dz \to \Km \pip, \Km \Kp, \Km \pip \piz, \Km \pip \pim \pip$, $\KS \pip \pim$;
$\Dp \to \Km \pip \pip, \Km \pip \pip \piz, \KS \pip, \KS \pip \pip \pim, \KS \pip \piz$, 
$\KS \Kp,$ with $\KS \to \pi^+ \pi^-$.  
For $D^*$ candidates, the decays $D^{*+}\to D^0\pi^+,D^+\pi^0$, and 
$D^{*0}\to D^0 \pi^0,D^0\gamma$ are used.

In events with more than one reconstructed \BB\ pair, we select the candidate with the lowest 
value of $\eextra$,  defined as the sum of the energies  of all photon candidates not associated with the 
reconstructed \BB\ pair. We further reject combinatorial background and normalization events 
by  requiring $q^2= (p_B - p_{D^{(*)}})^2>4\gev^2$ and $|\boldsymbol{p}_{\rm miss}|>200\mev$, where
 $|\boldsymbol{p}_{\rm miss}|$ is the missing momentum in the c.m. frame.

We divide the $\Btag\ds\ell$ candidates that satisfy the previous requirements into eight subsamples: 
four $\ds\ell$ samples, one for each of the
types of charm meson reconstructed ($D^0\ell$, $D^{*0}\ell$, $D^+\ell$, and $D^{*+}\ell$), and four
\dspizl control samples with the same selection plus an additional \piz.
These control samples constrain the poorly understood $\Bb\to\dss(\ell/\tau)\nu$ background (where
 \dss~refers to charm resonances heavier than the \Dstar ground state mesons),
 which enters the $\ds\ell$ sample 
predominantly when the \piz from $\dss\to\ds\piz$ decays is not detected.
The \dspizl samples have a very large continuum background, so we restrict this sample to events with
$|\cos \Delta \theta_{\rm thrust}|<0.8$, where $\Delta \theta_{\rm thrust}$ is the angle between 
the thrust axes of the \Btag and of the rest of the event.

We improve the separation between well-reconstructed events (signal and normalization)
and the various backgrounds by using boosted decision tree (BDT) selectors 
~\cite{BDT}.  
For each of the four $\ds\ell$ samples, we train a BDT to select 
signal and normalization events and reject $D^{**}\ell \nub$ background and charge cross-feed, 
defined as $\ds\elltau\nu$ decays reconstructed with the wrong charge. 
Each BDT selector relies on the simulated distributions of the following variables: 
(1) $\eextra$; 
(2) $\Delta E$; 
(3) the reconstructed mass of the signal $D$ meson; 
(4) the mass difference for the reconstructed signal $D^*$: $\Delta m= m(D\pi) - m(D)$; 
(5) the reconstructed mass of the seed meson of the \Btag; 
(6) the mass difference for a $D^*$ originating from the \Btag, $\Delta m_{\rm tag}= m(D_{\rm tag}\pi) - 
m(D_{\rm tag})$;
(7) the charged particle multiplicity of the \Btag candidate; and
(8) $\cos \Delta \theta_{\rm thrust}$. 
For the \dspizl samples, we use similar BDT selectors that are trained to reject continuum, 
$\ds(\ell/\tau)\nu$, and other \BB events.
After the BDT requirements are applied, the fraction of events attributed to signal in the 
$\mmiss>1.5\gev^2$ region, which excludes most of the
normalization decays, increases from 2\% to 39\%.
The remaining background is composed of normalization events (10\%), 
continuum (19\%), \Dsslnu events (13\%), and other 
\BB events (19\%), primarily from $B\to\ds D^{(*)+}_s$ decays with $\Ds \to \taup\nut$.

As described below, the fit procedure relies on the Monte Carlo (MC) simulation 
\cite{Sjostrand:1993yb, Lange:2001uf, Agostinelli:2002hh} of the 
two-dimensional \mmiss--\pstarl 
spectra of the different signal and background contributions.
For semileptonic decays, we
parameterize the hadronic matrix elements of the signal and normalization decays by using heavy-quark
effective theory-based
form factors (FFs)  \cite{Caprini:1997mu}. For low-mass leptons, there is effectively one FF 
for $\Bb \to D \ell^- \nub_{\ell}$, whereas there are three FFs for $\Bb \to D^* \ell^- \nub_{\ell}$ decays, 
all of which have been measured with good precision~\cite{Amhis:2012bh}.
For heavy leptons, each of these decays depends on an additional FF which can be calculated 
by using heavy-quark 
symmetry relations or lattice QCD. We use the calculations in Ref.~\cite{Tanaka:2010se} for \BDtaunu
and in Ref.~\cite{Fajfer:2012vx} for \BDstaunu.
For the \Dsslnu background, we consider in the nominal fit only the four $L=1$ states that have been  
measured~\cite{Aubert:2008ea}. We simulate these decays by using the Leibovich-Ligeti-Stewart-Wise 
calculation \cite{Leibovich:1997em}.

We validate and, when appropriate,  correct the simulations by using three data control samples
selected by one of the following criteria: $\eextra>0.5\gev$ \footnote{We do not place explicit 
requirements on \eextra, but the BDTs reject all events with $\eextra>0.4\gev$.}, $q^2\leq4\gev^2$, 
or $5.20<\mes<5.26\gev$.
We use off-peak data to correct the efficiency and the \pstarl spectrum of 
simulated continuum events.  After this correction, the \mmiss and 
\pstarl distributions of the background and normalization events agree very well with the simulation.
However, we find that small differences in the \eextra spectrum and other BDT input distributions 
result in a 5\%--10\% efficiency difference between selected data and MC samples.
We correct the continuum and \BB backgrounds by using the $5.20<\mes<5.26\gev$ control sample. 
The same correction, with larger uncertainties, is applied to \Dsslnu events, since
their simulated \eextra spectrum is very similar.  

We extract the signal and normalization yields from an extended, unbinned maximum-likelihood fit 
to two-dimensional \mmiss--\pstarl distributions.
The fit is performed simultaneously to the four $\ds\ell$ samples and the four \dspizl samples. 
The distribution of each $\ds\ell$ sample is described as the sum of eight contributions: 
$D\tau\nu$, $\Dstar\tau\nu$, $D\ell\nu$, $\Dstar\ell\nu$, $\dss(\ell/\tau)\nu$, charge cross-feed, other \BB, 
and continuum. The yields for the last three backgrounds are fixed in the fit to the expected values. 
A large fraction of $B\to D^{*}\ell\nu$ decays (for $B=\Bz$ or \Bp) is reconstructed in the $D\ell$ samples 
(feed-down).
We leave those two contributions free in the fit and use the fitted yields to estimate the feed-down 
rate of $B\to \Dstar\tau\nu$ decays.
Since $B\to D(\ell/\tau)\nu$ decays contributing to the  $\Dstar\ell$ samples are rare, their
rate is fixed to the expected value. 

The four \dspizl samples are described by six contributions: 
The $\ds\tau\nu$ and $\ds\ell\nu$ yields are combined but otherwise
the same contributions that describe the $\ds\ell$ samples are employed.
The four \Dsslnu yields in the control samples are free in the fit, but their ratios to the 
corresponding \Dsslnu yields in the $\ds\ell$ samples are constrained to the expected values.

The fit relies on $8\times4+6\times4=56$ probability density functions (PDFs), which are determined 
from MC samples of continuum and \BB\ events equivalent to 2 and 9 times the size of the 
data sample, respectively. The two-dimensional \mmiss--\pstarl
distributions are described by using smooth nonparametric kernel estimators \cite{Cranmer:2000du}.
The fit is iterated to update some of the parameters that depend on the normalization yields, most 
importantly the rate of signal feed-down. This process converges after the first iteration.
We performed MC studies to verify that neither the fit procedure nor the PDFs produced
significant biases in the results.

\begin{figure}[tb!]\begin{center}
\includegraphics[width=3.18in]{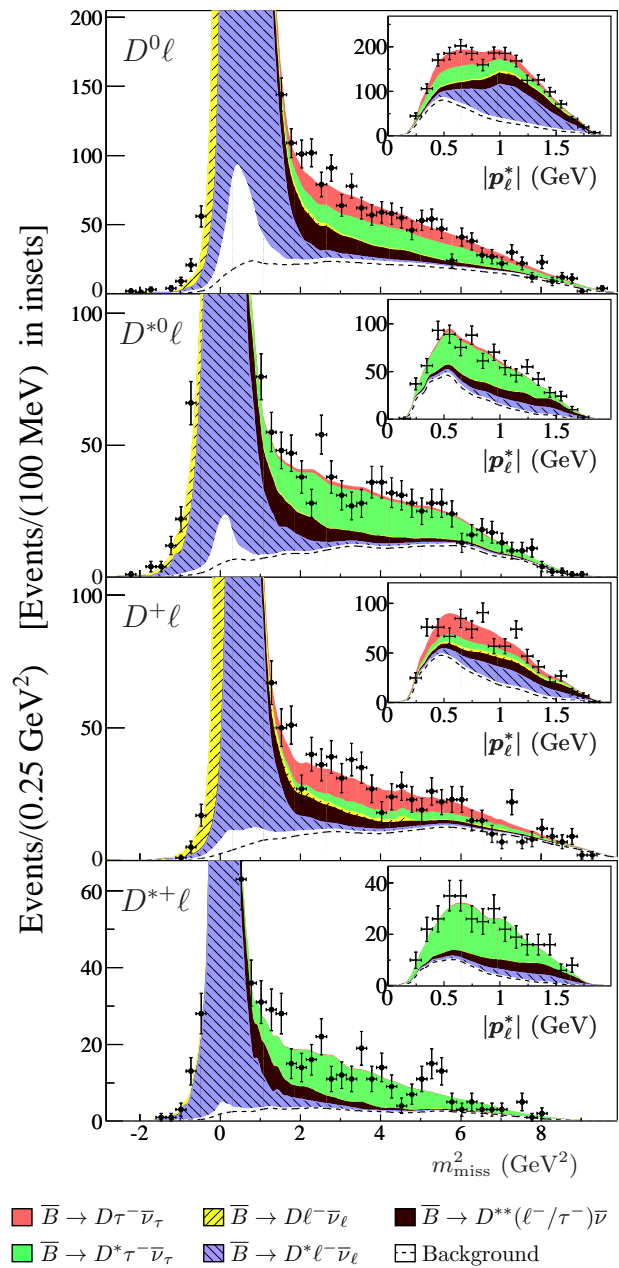} 
\caption {(Color online) Comparison of the data and the fit projections 
for the four $\ds\ell$ samples. 
The insets show the \pstarl projections for $\mmiss>1\gev^2$, which excludes most of the 
normalization modes. In the background component, the
region above the dashed line corresponds to charge cross-feed, and the region
below corresponds to continuum and \BB.}\label{fig:PRL_Fit}
\end{center}\end{figure}

Figure \ref{fig:PRL_Fit} shows the \mmiss and \pstarl projections of the fit to the four $\ds\ell$ samples. 
The fit describes the data well and the observed differences are consistent with the statistical and systematic  
uncertainties  on the signal PDFs and background distributions.

We extract the branching fraction ratios as  
$\RDx =(N_{\rm sig}/N_{\rm norm})/(\eps_{\rm sig}/\eps_{\rm norm})$,
where $N_{\text{sig}}$  and $N_{\text{norm}}$ refer to the number of signal and normalization events,
respectively, and $\eps_{\text{sig}}/\eps_{\text{norm}}$ is the ratio of their efficiencies derived from simulations. 
Table \ref{tab:FinalResults} shows the results of the fits for the four individual samples as well as 
an additional fit in which we impose the isospin relations 
$\RDz=\RDp\equiv\RD$ and $\RDstarz=\RDstarp\equiv\RDs$.
The statistical correlations are $-0.59$ for \RDz and \RDstarz,
$-0.23$ for \RDp and \RDstarp, and $-0.45$ for \RD\ and \RDs.
We have verified that the values of \RDx\ from fits to samples corresponding to different run 
periods are consistent. We repeated the analysis varying the selection criteria over a wide range
corresponding to changes in the  signal-to-background ratios between 0.3 and 1.3,
and also arrive at consistent values of \RDx.

\begin{table*} [hbt] 
\caption{Results of the isospin-unconstrained (top four rows) and isospin-constrained fits 
(last two rows). The columns show the 
signal and normalization yields, the ratio of their efficiencies, \RDx, branching fractions,  
and $\Sigma_{\rm stat}$ and $\Sigma_{\rm tot}$,
the statistical and total significances, respectively. 
Where two uncertainties are given, the first is statistical and the second is systematic. The branching fractions
${\cal B}(\BDxtaunu)$ are calculated as $\RDx\times{\cal B}(\BDxlnu)$, using the average 
\BDxlnu branching fractions measured by \babar~\cite{PhysRevLett.104.011802, PhysRevD.79.012002,
PhysRevD.77.032002}.
The stated branching fractions for the isospin-constrained fit refer to \Bm\ decays.} 
\label{tab:FinalResults} \vspace{0.1in}
\begin{tabular}{ll r @{ $\pm$ } l r @{ $\pm$ } l r @{ $\pm$ } l r @{ $\pm$ } l @{ $\pm$ } l r @{ $\pm$ } l @{ $\pm$ } l rr} \hline\hline
\multicolumn{2}{l}{Decay} &  \multicolumn{2}{c}{$N_\mathrm{sig}$} &  \multicolumn{2}{c}{$N_\mathrm{norm}$} 
&  \multicolumn{2}{c}{$\eps_{\rm sig}/\eps_{\rm norm} $} 
&  \multicolumn{3}{c}{\RDx} &  \multicolumn{3}{c}{${\cal B}(B\to\ds\tau\nu)\,(\%)$} 
&$\Sigma_{\text{stat}}$	& $\Sigma_{\text{tot}}$  \\ \hline
$\Bm\!\to$&$\Dz\taum\nutb$	&\hspace{0.05in}  314	& 60&\hspace{0.1in} 1995	& 55&\hspace{0.1in} 0.367	&0.011\hspace{0.1in}
				& 0.429	& 0.082	& 0.052\hspace{0.1in}	
				& 0.99	& 0.19\hspace{0.1in}& 0.13	& 5.5 &\hspace{0.15in}4.7 \\
$\Bm\!\to$&$\Dstarz\taum\nutb$& 639	& 62		& 8766	& 104	& 0.227	& 0.004	& 0.322	& 0.032	& 0.022	
				& 1.71	& 0.17	& 0.13	& 11.3 	&9.4 \\
$\Bzb\to$&$\Dp\taum\nutb$	& 177	& 31		& 986	& 35		& 0.384	& 0.014	& 0.469	& 0.084	& 0.053	
				& 1.01	& 0.18	& 0.12	& 6.1 	&5.2 \\
$\Bzb\to$&$\Dstarp\taum\nutb$& 245	& 27		& 3186	& 61		& 0.217	& 0.005	& 0.355	& 0.039	& 0.021	
				& 1.74	& 0.19	& 0.12	& 11.6 	&10.4 \\ \hline
$\Bb\;\to$&$ D\taum\nutb$	& 489	& 63		& 2981	& 65		& 0.372	& 0.010	& 0.440	& 0.058	& 0.042	
				& 1.02	& 0.13	& 0.11	& 8.4		&6.8 \\
$\Bb\;\to$&$\Dstar\taum\nutb$	& 888	& 63		& 11953	& 122	& 0.224	& 0.004	& 0.332	& 0.024	& 0.018	
				& 1.76	& 0.13	& 0.12	& 16.4 	&13.2 \\ \hline\hline
\end{tabular}
\end{table*}

The largest systematic uncertainty affecting the fit results is due to the poorly understood \BDssltnu
background. The PDFs that describe this contribution 
are impacted by the uncertainty on the branching fractions of the four \BDsslnu decays,
the relative $\piz/\pipm$ efficiency, and the branching fraction ratio of \BDsstaunu
to \BDsslnu decays. These effects contribute to an uncertainty of 2.1\% on \RD\ and 1.8\% on \RDs.
We also repeated the fit including an additional floating component with the distributions of
$B\to\ds\eta\ell\nu$, nonresonant $B\to\ds\pi(\pi)\ell\nu$, and $B\to\dss(\to\ds\pi\pi)\ell\nu$ decays. 
The \BDssltnu background is tightly constrained by the \dspizl samples,
and, as a result, all these fits yield similar values for \RDx. We assign the observed variation
as a systematic uncertainty, 2.1\% for \RD\ and 2.6\% for \RDs.

We also account for the impact of the uncertainties described above on the relative efficiency 
of the \BDssltnu contributions to the signal and \dspizl samples. In addition, the BDT selection
introduces an uncertainty that we estimate as 100\% of the efficiency correction that we determined 
from control samples. These effects result in uncertainties of  5.0\% and 2.0\% on \RD\ and
\RDs, respectively.

The largest remaining uncertainties are due to the continuum and \BB backgrounds [4.9\% on \RD\ and 
2.7\% on \RDs],
and the PDFs for the signal and normalization decays (4.3\% and 2.1\%). 
The uncertainties in the efficiency ratios $\eps_{\rm sig}/\eps_{\rm norm}$
are 2.6\% and 1.6\%; they do not affect the significance of the signal and are dominated by the limited 
size of the MC samples. Uncertainties due to the FFs, particle identification, final-state radiation, 
soft-pion reconstruction, and others related to the detector performance largely
cancel in the ratio, contributing only about 1\%.
The individual systematic uncertainties are added in quadrature to define the total systematic
uncertainty, reported in Table \ref{tab:FinalResults}.

There is a positive correlation between some of the systematic uncertainties on \RD\ and \RDs, 
and, as a result
the correlation of the total uncertainties is reduced to  $-0.48$ for \RDz and \RDstarz,
to $-0.15$ for \RDp and \RDstarp, and to $-0.27$ for \RD\ and \RDs.

The statistical significance of the signal is determined as $\Sigma_{\rm stat}=\sqrt{2\Delta(\text{ln}{\cal L})}$,
where $\Delta(\text{ln}{\cal L})$ is the change in the log-likelihood between the nominal fit and the 
no-signal hypothesis. The statistical and dominant systematic uncertainties are Gaussian.
We estimate the overall significance as  
$\Sigma_{\rm tot}=\Sigma_{\rm stat}\times \sigma_{\rm stat}/\sqrt{\sigma_{\rm stat}^2+\sigma_{\rm syst}^{*2}}$,
where $\sigma_{\rm stat}$ is the statistical uncertainty and $\sigma^*_{\rm syst}$ is the 
total systematic uncertainty affecting the fit.  
The significance of the \BDtaunu signal is $6.8\sigma$, the first such measurement exceeding $5\sigma$.

To compare the measured \RDx\ with the SM predictions we have updated the 
calculations in Refs.~\cite{Kamenik:2008tj,Fajfer:2012vx} taking into account recent FF 
measurements.
Averaged over electrons and muons, we
find $\RD_{\rm SM}=0.297\pm0.017$ and $\RDs_{\rm SM}=0.252\pm0.003$.
At this level of precision, additional uncertainties 
could contribute~\cite{Fajfer:2012vx}, but the experimental uncertainties are expected to dominate.

\begin{figure}[tb!]
\psfrag{R\(D\)}[bl]{\RD}
\psfrag{R\(D*\)}[bl]{\RDs}
\psfrag{t}[Br]{\tBmH (GeV$^{-1}$)}
\includegraphics[width=3.4in]{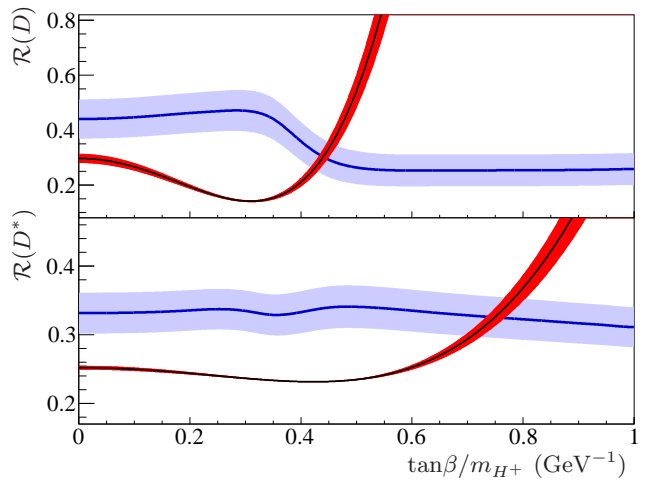}
\caption {(Color online) Comparison of the results of this analysis (light gray, blue)
with predictions that include a charged Higgs boson of type II 2HDM (dark gray, red).  
The SM corresponds to $\tanB/\mH=0$. }
\label{fig:PRL_Higgs}
\end{figure}

Our measurements exceed the SM predictions for \RD\ and \RDs\ by $2.0\sigma$ and $2.7\sigma$, 
respectively. The 
combination of these results, including their $-0.27$ correlation, yields $\chi^2=14.6$ for 2 degrees of 
freedom, corresponding to a $p$ value of $6.9\times10^{-4}$.
Thus,  the possibility of both the measured 
\RD\ and \RDs\ agreeing with the SM predictions is excluded at the $3.4\sigma$ level \footnote{Recent
calculations \cite{Nierste:2008qe,Tanaka:2010se,Bailey:2012jg} have found values of 
$\RD_{\rm SM}$\ that somewhat exceed our estimation. With the largest of those values, the
significance of the excess decreases to $3.2\sigma$.}.

Figure \ref{fig:PRL_Higgs} shows the effect that a charged Higgs boson of the type II 2HDM 
\cite{Tanaka:2010se,Barger:1989fj} would have
on \RD\ and \RDs\ in terms of the ratio of the vacuum expectation values $\tanB\equiv v_2/v_1$, and the
mass of the charged Higgs \mH. 
We estimate the effect of the 2HDM on our measurements by reweighting the simulated events
at the matrix element level for 20 values of \tBmH over the $[0.05,1]\gev^{-1}$ range. We then repeat
the fit with updated PDF shapes and $\eps_{\rm sig}/\eps_{\rm norm} $ values. The increase in the uncertainty 
on the PDFs and the efficiency ratio is estimated for each value of \tBmH. The other sources of systematic uncertainty
are kept constant in relative terms.

The measured values of \RD\ and \RDs\ match the predictions of this particular Higgs model
for  $\tanB/\mH=0.44\pm0.02\gev^{-1}$ and $\tanB/\mH=0.75\pm0.04\gev^{-1}$, respectively.
However, the combination of \RD\ and \RDs\ excludes the type II 2HDM charged Higgs boson with a 
99.8\% confidence level for any value of \tBmH.
This calculation is valid only for values of \mH greater than about $10\gev$~\cite{Tanaka:1994ay,Tanaka:2010se}. 
The region for $\mH\leq 10\gev$ has already been excluded by $B\to X_s\gamma$ 
measurements \cite{Misiak:2006zs}, and, therefore, the type II 2HDM is excluded
in the full  \tanB--\mH parameter space.

In summary, we have measured the \BDtaunu and \BDstaunu decays relative to the decays
to light leptons \BDxlnu. We find
\begin{eqnarray}
\RD &=& 0.440 \pm 0.058 \pm 0.042 \nonumber\\
\RDs &=& 0.332 \pm 0.024 \pm 0.018~. \nonumber
\end{eqnarray}
These results supersede the previous \babar\ results and have significantly reduced uncertainties. 
The measured values are compatible with those measured by the Belle Collaboration
\cite{Matyja:2007kt,Adachi:2009qg,Bozek:2010xy}. 

The results presented here disagree with the SM at the $3.4\sigma$ level, which, 
together with the measurements by the Belle Collaboration, could be an indication of 
new physics processes affecting \BDxtaunu decays.
However, our results are not compatible with the widely discussed type II 2HDM for any value of
\tanB and \mH.

\begin{acknowledgments}
We acknowledge M. Mazur for his help throughout the analysis, and
S. Westhoff, S. Fajfer, J. Kamenik, and I. Ni\v{s}and\v{z}i\'{c} for their help with the calculation 
of the charged Higgs contributions.
We are grateful for the excellent luminosity and machine conditions
provided by our \pep2\ colleagues, 
and for the substantial dedicated effort from
the computing organizations that support \babar.
The collaborating institutions thank 
SLAC for its support and kind hospitality. 
This work is supported by
DOE and NSF (USA), NSERC (Canada),
CEA and CNRS-IN2P3
(France), BMBF and DFG
(Germany), INFN (Italy),
FOM (The Netherlands),
NFR (Norway), MES (Russia),
MICIIN (Spain), and STFC (United Kingdom). 
Individuals have received support from the
Marie Curie EIF (European Union)
and the A.~P.~Sloan Foundation (USA).
\end{acknowledgments}

\bibliography{paper}

\begin{thebibliography}{35}
\bibitem [{con()}]{convention}%
  \BibitemOpen
  \href@noop {} {}\bibinfo {note} {Throughout this letter, $\ell$ refers only
  to the light leptons $e$ and $\mu$, $D^{(*)}$ refers to a $D$ or a $D^*$
  meson, and charge-conjugate decay modes are implied}\BibitemShut {NoStop}%
\end{thebibliography}

\end{document}